\documentclass[aps,prb,amssymb, amsmath, twocolumn, superscriptaddress]{revtex4-2}

\usepackage{amssymb}
\usepackage{amsmath}
\usepackage{bbm}
\usepackage{bm}
\usepackage{braket}
\usepackage{graphicx}
\usepackage[citecolor=blue,linkcolor=blue,urlcolor=blue,colorlinks=true]{hyperref}
\usepackage{latexsym}
\usepackage{lineno}
\usepackage{multirow}
\usepackage{xcolor}

\begin{document}
\title{Fractal Surface States in Three-Dimensional Topological Quasicrystals}
\author{Zhu-Guang Chen}
\affiliation{Zhejiang Institute of Modern Physics, Zhejiang University, Hangzhou 310027, China}
\author{Cunzhong Lou}
\affiliation{Zhejiang Institute of Modern Physics, Zhejiang University, Hangzhou 310027, China}
\author{Kaige Hu}
\email{hukaige@gdut.edu.cn}
\affiliation{Department of Physics, School of Physics and Optoelectronic Engineering, Guangdong University of Technology, Guangzhou 510006, China}
\author{Lih-King Lim}
\email{lihking@zju.edu.cn}
\affiliation{Zhejiang Institute of Modern Physics, Zhejiang University, Hangzhou 310027, China}
\date{\today}

\begin{abstract}
We study topological states of matter in quasicrystals, which do not rely on crystalline orders. In the absence of a bandstructure description and spin-orbit coupling, we show that a three-dimensional quasicrystal can nevertheless form a topological insulator. It relies on a combination of noncrystallographic rotational symmetry of quasicrystals and electronic orbital space symmetry, which is the quasicrystalline counterpart of topological crystalline insulator. The resulting topological state obeys a non-trivial twisted bulk-boundary correspondence and lacks a good metallic surface. The topological surface states, localized on the top and bottom planes respecting the quasicrystalline symmetry, exhibit a new kind of multifractality with probability density concentrates mostly on high symmetry patches. They form a near-degenerate manifold of `immobile' states whose number scales proportionally with the macroscopic sample size. This can open the door to a novel platform for topological surface physics distinct from the crystalline counterpart. 
\end{abstract}

\maketitle
\section{Introduction}
The modern classification of topological states according to symmetry classes and dimensionality provides an organizational principle in the search of new states of matter \cite{Hasan2010rmp, Qi2011rmp, Moessner2021}. One key ingredient is the bandstructure theory afforded by crystalline environment. Recently, a huge effort is to enlarge the search of topological states for systems without translational invariance. They include quasicrystals \cite{Kraus2012prl, Fulga2016prl, Bandres2016prx, Huang2018prl, Varjas2019prl, Chen2020prl, Andr2023prb} and amorphous topological matter \cite{Agarwala2017prl, Xiao2017prb, Mansha2017prb, Mitchell2018natphy, Poyh2018natcom, Xu2019prl, Marsal2020pnas, Sahlberg2020prr, Agarwala2020prr, Xu2021prl, Wang2022prl}, where despite the lack of crystallinity, the resulting states bear all the hallmark of topological states. These systems are not small disorder perturbations from the crystalline limit, and thus their existence are examples showing the generality of topological concepts. 

In the absence of crystallinity, it brings about two relevant questions. First, it calls for new methodology to identify the topological invariant since the well-established momentum-space topological invariants are inapplicable \cite{Kitaev2006, Bianco2011prb, Loring2011epl, Prodan2011}. Second, a more salient feature is the noncrystalline surface state property, which is responsible for topological responses \cite{Bandres2016prx, Mansha2017prb, Huang2018prl, Agarwala2020prr}. For the former, much progress has been made in terms of real space formulation of topological invariants \cite{Kitaev2006, Bianco2011prb, Loring2011epl, Prodan2011}, generally called topological markers. Beside the Bott index for two-dimensional quasicrystals \cite{Huang2018prl} and amorphous topological systems \cite{Agarwala2017prl, Mitchell2018natphy} (of the quantum spin Hall and integer quantum Hall classes, respectively), a topological marker for three-dimensional noncrystalline systems, the analogue of strong topological insulators, has been recently established \cite{Hannukainen2022prl, Chen2023prb}. 

As for the surface states, the absence of Bloch-wave description that underpins nontrivial surface band crossing \cite{Hasan2010rmp} offers a new platform to test topological robustness. Works have so far been focussed on effectively low-dimensional edges, for example in the perimeter of a two-dimensional \cite{Chen2020prl,Agarwala2020prr} or three-dimensional higher-order topological insulators \cite{Agarwala2020prr, Xu2021prl, Mao2024prb}, realized in both quasicrystal and amorphous matter. Symmetry protection continues to support a robust, scattering-free transport giving rise to quantized conductance in spite of their irregular edges. In higher dimensional consideration, on the other hand, it remains a largely unexplored topic \cite{Poyh2018natcom, WangC2022prl} which is due in part to the scarcity of higher dimensional noncrystalline topological systems.   

\begin{figure}[b]
    \centering
    \includegraphics[width=0.48\textwidth]{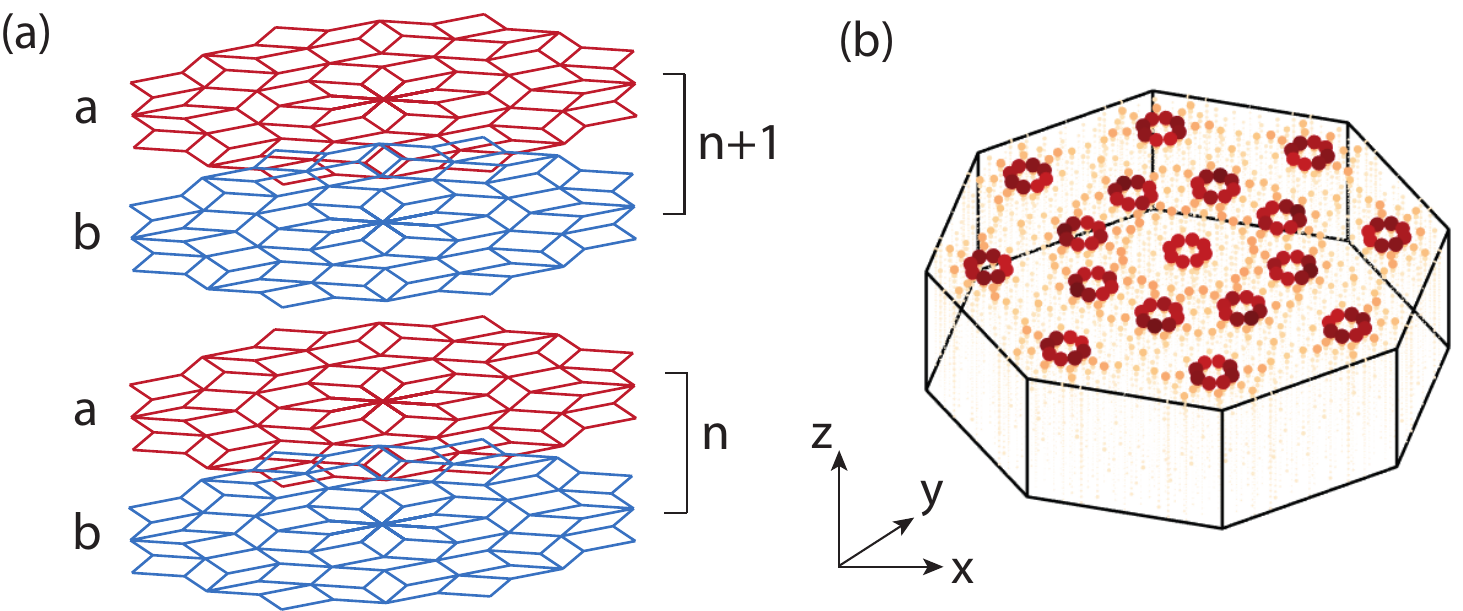}
    \caption{Three-dimensional topological quasicrystal. (a) Stack of Ammann-Beenker quasicrystal bilayers with tight-binding orbital electrons hopping between vertices. (b) Probability density distribution of topological surface states covering A-patches with multifractality on the quasicrystal top-plane.}
    \label{fig1} 
\end{figure}

In this work, we consider a three-dimensional realization of topological quasicrystalline insulator, utilizing the electron orbital states with noncrystallographic symmetry of quasicrystal \cite{Senechal1996quasicrystals, baake2013aperiodic1}. We take, as an example, a stack of an eight-fold rotational symmetric Ammann-Beenker quasicrystal bilayers forming a three-dimensional structure, see Fig.~\ref{fig1}(a). With tight-binding $p$-band spinless electrons, the result is a noncrystalline counterpart of topological crystalline insulator \cite{Fu2011prl, Slager2013nphys}, which \textit{stricto sensu} exists only for three-dimensional system and does not require spin-orbit coupling.  As the latter belongs to the fragile topological class \cite{Po2018prl, Cano2018prl, Alexandradinata2020prb}, and moreover, by subjecting to noncrystallinity we can employ a generalized twisted bulk-boundary correspondence \cite{Song2020sicence} to unveil the underlying nontrivial topology. The topological surface states, localized on the quasicrystal top-(or bottom-)plane, see Fig.~\ref{fig1}(b), exhibit multifractality \cite{Kohmoto1987prb, Repetowicz1998prb, Jagannathan2017prb, Oktel2021prb, Evers2008rmp, Jagannathan2021rmp} with probability weights concentrating only on high-symmetry patches of the lattice - they thereby inherit planar quasicrystal eigenstate characteristics that are neither Bloch-type surface modes \cite{Fu2011prl} nor localized corner states \cite{Benalcazar2017prb,Song2017prl,Varjas2019prl}. These characteristics result in `immobile' states populating a near-degenerate energy manifold, with their number - macroscopic in number - determined by high-symmetry patches of the quasicrystal plane. With interactions or superconductivity, one can envisage a potentially novel platform for topological surface physics \cite{Ye2020prl, Nature2023}.

\section{Model Hamiltonian}
\begin{figure}[t]
    \centering
    \includegraphics[width=0.475\textwidth]{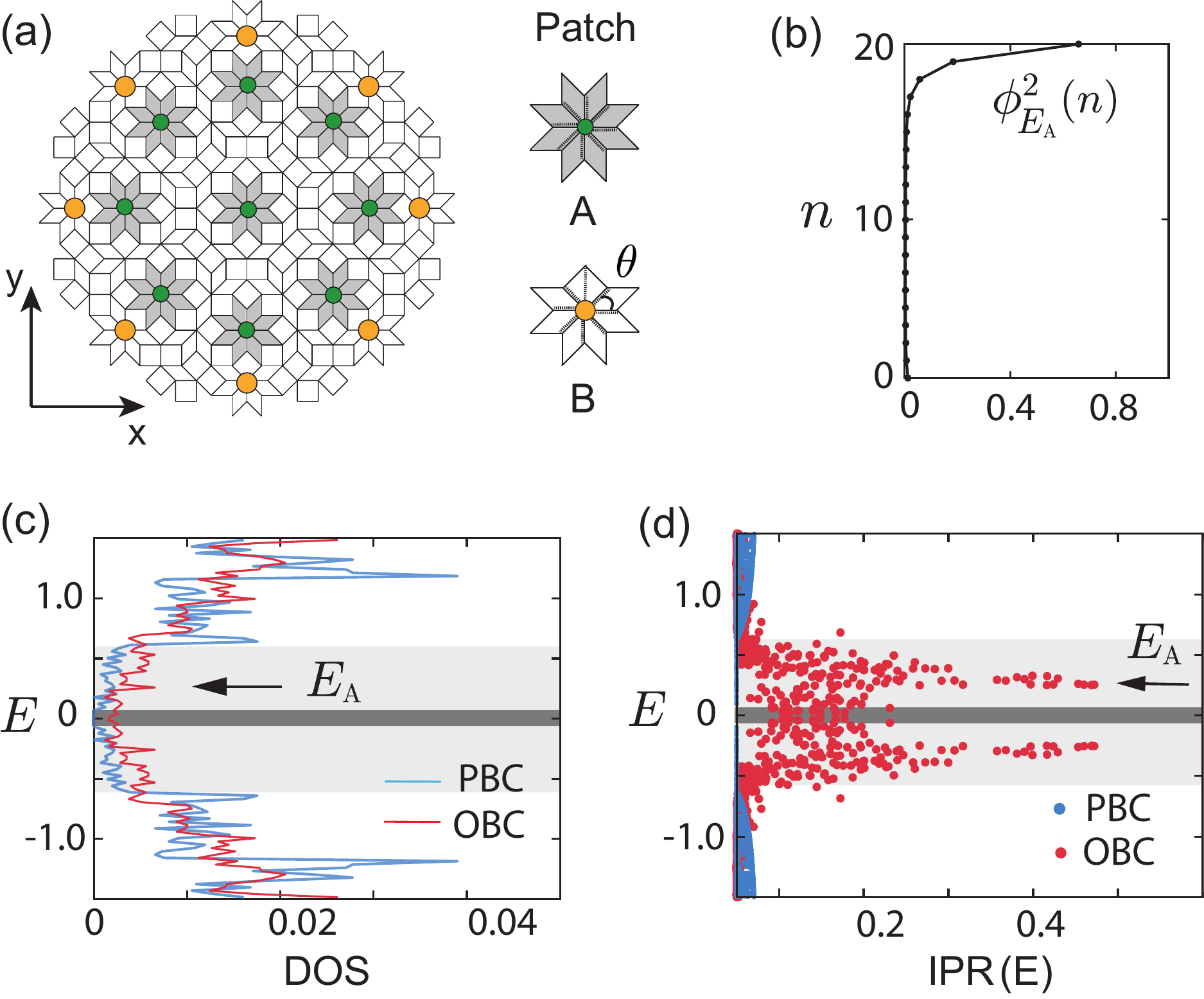}
    \caption{Ammann-Beenker tiling and spectral properties of Hamiltonian (\ref{eq1}). (a) Center sites of A- and B-patches (in green and yellow, respectively) and hopping paths (dotted lines). (b) Probability density distribution in $z$-direction for eigenstates with energy $E_\textrm{A}=0.25$.  (c)  Density-of-states (DOS) in the center of the energy spectrum. (d) Inverse participation ratio (see text). Blue (red) with periodic (open) boundary condition in the $z$-direction. Tight-binding parameters: $t_1=1.7, t_3=0.5, t_z=2.5, t_z'=2$, \textrm{ with $689\times 20\times 2$ sites (setting energy unit $t_2=1$ throughout the work).}}
    \label{fig2} 
\end{figure}

The three-dimensional model is built by stacking two-dimensional Ammann-Beenker quasicrystal bilayers, with atoms positioned at vertices and stacked directly above each other. We then consider spinless electrons from two orbitals, ${p_x}$, ${p_y}$, on each site, hopping to neighboring sites within a $\sqrt{2}$-distance of the tiling edge length, see Fig.~\ref{fig2}(a) and Appendix \ref{A1}. The Hamiltonian is given by
\begin{equation} 
H=\sum_{\mathbf{R}_i,\mathbf{R}_j}\sum_{\alpha,\beta}\, t_{ij,\alpha\beta}|\mathbf{R}_i,\alpha\rangle \langle \mathbf{R}_j,\beta|
\label{eq1}
\end{equation}
with electron state $|\mathbf{R}_i,\alpha\rangle$ at position $\mathbf{R}_i=\{n,\sigma,\mathbf{r}_i\}$ denoting the $n$-th bilayer of the $\sigma=a,b$ plane with coordinate $\mathbf{r}_i=(x_i,y_i)$, and $\alpha,\beta=p_x,p_y$ denoting the orbital state. By representing the coupling of $p_{x,y}$ orbitals in a pseudospin space spanned by the eigenstates of Pauli matrix $\tau_z$, the in-plane hopping matrix can be written as
\begin{equation} 
 t_{ij}=\pm t(|\mathbf{d}_{ij}|)\,\left(
\begin{array}{cc}
	\cos^2\theta &  \frac{1}{2}\sin2\theta  \\
	 \frac{1}{2}\sin2\theta & \sin^2\theta
\end{array}\right)
\label{eq2}
\end{equation}
with the angle $\theta$ for the displacement vector between any two sites $\mathbf{d}_{ij}=\mathbf{r}_j-\mathbf{r}_i$ with the $x$-direction [Fig.~\ref{fig2}(a)], and $\pm t(|\mathbf{d}_{ij}|)$ are hopping strength ($\pm$ signature for $a/b$ layer, respectively). Applying the $\sqrt{2}$-distance tight-binding parameters, there are three possible in-plane hopping strengths $t_{1,2,3}$ in descending order for three distances $|\mathbf{d}_{ij}|\leq \sqrt{2}$ in unit of the tiling edge length. For intra-bilayer hoppings (i.e., $\sigma$-changing with the same $n$), we assume they are orbital independent, with overall amplitudes reduced by a factor one-half with constant signature, and the hopping matrix replaced by an identity matrix. Furthermore, hopping along the $z$-direction between neighboring planes is dimerized with amplitudes $t_z, t'_z$. The system exhibits a combined $C_2$ symmetry based on the product of the $C_8$ symmetry generator of the orbital and coordinate spaces. The unusual eight-fold rotational generator is due to the underlying quasicrystal structure of Ammann-Beenker, which can be generalized to other quasicrytalline rotational orders such as $C_5$ order of the Penrose tiling (see below) \cite{Senechal1996quasicrystals, baake2013aperiodic1}. They are noncrystalline generalizations of Fu's model with crystalline rotational order of $C_4$ or $C_6$ symmetry \cite{Fu2011prl, Slager2013nphys, Alexandradinata2016prb}.  

\section{Energy spectrum, symmetry of surface states}
We numerically obtain the energy spectrum of Hamiltonian (\ref{eq1}) for a finite patch in the $xy$ plane and with two different kinds of termination in the $z$-direction, either an open or a periodic boundary condition. Because the system lacks a bandstructure description, we further compute eigenstate-by-eigenstate $\psi_E(\mathbf{R}_i,\alpha)$ the inverse participation ratio $\textrm{IPR}(E)\equiv \sum_n |\phi_E(n)|^4$ with respect to the $n$-th bilayer basis. Here, $\phi_E(n)\equiv (\sum_{\sigma,\mathbf{r}_i,\alpha}|\psi_E|^2)^{1/2}$ is the normalized $n$-th bilayer eigenstate with energy $E$ and the IPR measures its degree of localization in the $z$-direction.

As a function of the inter-bilayer coupling $t'_z/t_z$ we establish a distinct state when the coupling is sufficiently strong. Starting with an ordinary quasicrystal insulator, the bulk gap closes and reopens as the coupling increases, accompanying with the presence of in-gap states with the open boundary condition [Fig.~\ref{fig2}(c)]. This is corroborated by the open-boundary IPR spectrum, where many states proliferated in the bulk gap region with $\textrm{IPR}>10^{-1}$ developing into two peaks at energies $\pm E_\textrm{A}$, respecting a particle-hole symmetry [Fig.~\ref{fig2}(d)]. Within the energy shell $E_\textrm{A}\pm\delta$, for variable $\delta$, the eigenstates give rise to the top surface probabilty density distribution in Fig.~\ref{fig1}(b) along with their localization in the $z$-direction in Fig.~\ref{fig2}(b). The particle-hole symmetric partners at $-E_\textrm{A}\pm\delta$ are, on the other hand, localized on the bottom surface. The energy range covering localized states [light shade in Fig.~\ref{fig2}(c, d)] extends beyond the bulk gap (in dark shade, respectively) is a result of mixtures of bulk and surface states, where the latter can be attributed to in-gap states in a projected bandstructure picture. It is, however, crucial to distinguish these surface states from three-dimensional quantum Hall systems, which are typically realized by stacking two-dimensional quantum Hall systems with metallic surface states in parallel to the stacking direction \cite{Halperin1987jpn}. Here, the localized states are realized only on planes displaying the quasicrystal rotational order, e.g., cut perpendicular to the rotational axis. 

In the following, we further resolve that there are states in the IPR peak exhibiting strictly two-fold degeneracy. They are the analog of symmetry protected surface band crossing of topological insulators with Bloch bands \cite{Fu2011prl, Hasan2010rmp}. We define a combined orbital-real space rotation by 
\begin{equation}   
\hat{U}_{ml}=\exp\left(\frac{-i\pi m\tau_{y}}{4}\right)\hat{R}_{\frac{\pi{l}}{4}},
\label{eq3}
\end{equation}
where the first factor perform a $m \pi/4$ rotation in the orbital space and the second factor is the ordinary rotation operator of angle $l \pi/4$ on the quasicrystal around the $z$-axis, with $m,l$ integers. It is a real operator. The symmetry is realized with $[H, \hat{U}_{ml}]=0$ when $m+l=4 \textrm{ (mod)}$, that is an effective $C_2$ symmetry of the tight-binding model. In fact, this symmetry holds also for the Hamiltonian $\mathcal{H}(k_z)$ with $k_z$ the Bloch momentum reciprocal to the $n$-th bilayer index in $H$, as well as the two-dimensional Hamiltonian $H_{\text{2D}}$ for each plane. Of particular relevance is the symmetry operator $\hat{U}_{ml}$ can have a maximum of eight eigenvalues with unit modulus, with three pairs of complex eigenvalues that are related by complex conjugation $1\pm i$, $\pm i$, $-1\pm i$. Since both the Hamiltonian $H$ (time-reversal invariant with spinless electrons) and $\hat{U}_{ml}$ are real, the degenerate energy eigenstates are guaranteed to be two-fold degenerate, distinguished by conjugate pair of complex eigenvalues for the $\hat{U}_{ml}$ operator. Focussing on eigenstates in the energy shell $E_\text{A}\pm\delta$ for small $\delta$, we identify three such pairs of degenerate states along with two additional isolated states (with eigenvalues $\pm 1$), forming a near-degenerate multiplet of states (Appendix \ref{A3}). 

Next, we inspect wavefunctions from the multiplet on the top surface, see Fig.~\ref{fig1}(b) with $\delta=0.03$. Evidently, they are organized in a regular pattern respecting the eight-fold rotational symmetry, with probability density concentrating only on the so-called A-patch of the Ammann-Beenker tiling - one of the seven vertex enviroments permissible in the quasicrystal with maximal edges, see Fig.~\ref{fig2}(a) and Appendix \ref{A2}. As the system size increases in the $xy$ plane respecting the rotational order, new surface states populating the energy shell $E_\text{A}\pm\delta$ appear only when a new generation of A-patch is enclosed (Appendix \ref{A2}). The quasicrystal plane thus supports `localized' surface modes that are neither Bloch-like, nor strictly localized, but follow the self-similar generation rule of Ammann-Beenker tiling. In fact, as we show below, they exhibit fractal structure.  

\section{Spectral flow and topological phase}
In the absence of translational invariance in the $xy$ plane, the usual characterization of momentum space topology is inapplicable. Morevover, known three-dimensional topological markers \cite{Hannukainen2022prl, Chen2023prb, Lee2023prb} fail to distinguish the phases as the topology stems from a global discrete rotational order. Here, we adapt the twisted bulk-boundary correspondence for fragile topology \cite{Song2020sicence}, to show that the appearance of surface states is associated with a topological phase transition. 

\begin{figure}[t]
    \centering
    \includegraphics[width=0.48\textwidth]{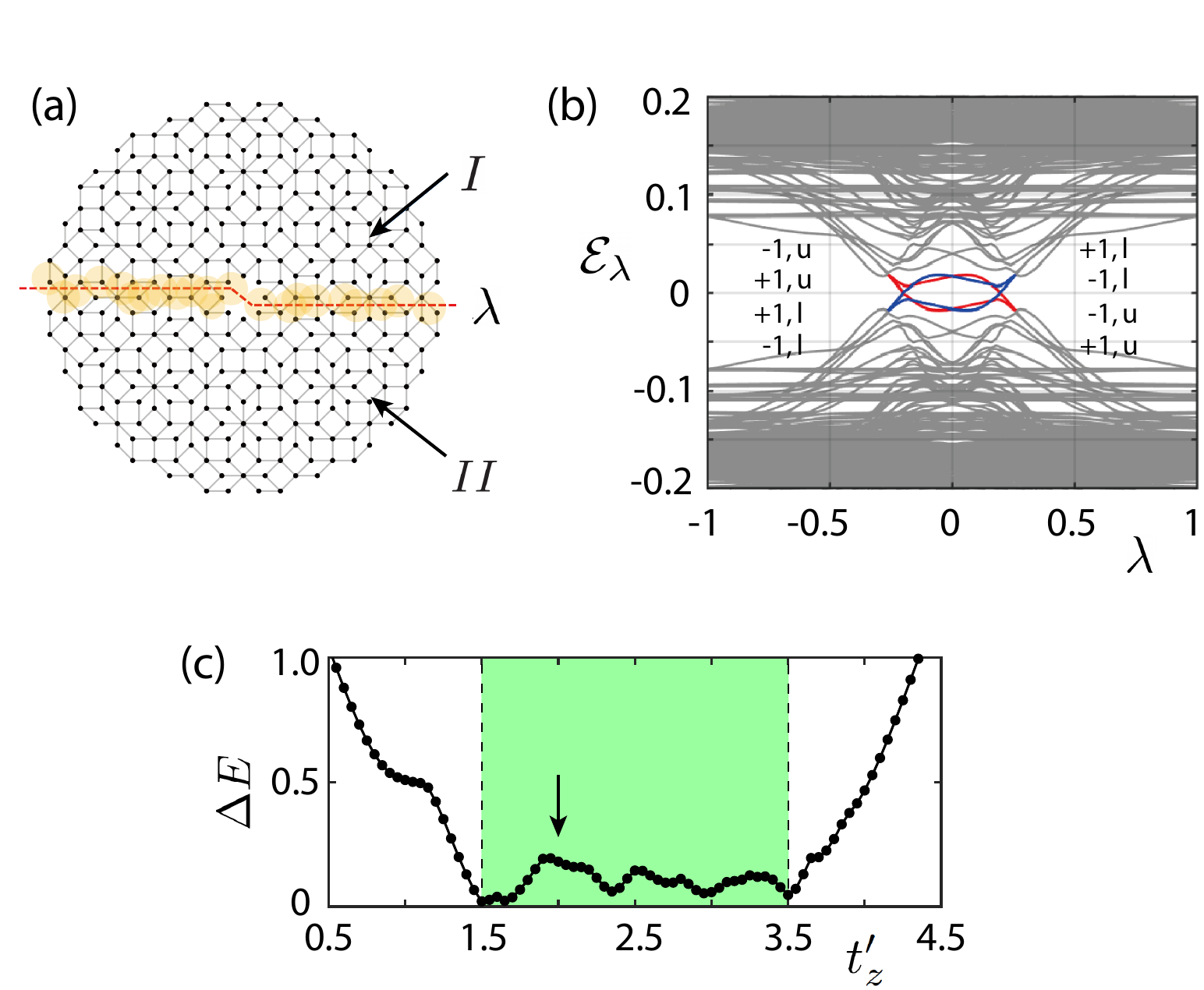}
        \caption{Generalized twisted boundary condition and topological phase diagram. (a) Lattice of $\mathcal{H}_{\lambda}(k_z)$ with $6080 \times 2$ sites divided into part I and II, and hopping strength crossing the red dashed line are multiplied by $\lambda$. Yellow shades are distribution of typical in-gap state from (b). (b) Spectral flow for $\mathcal{H}_{\lambda}(k_z=\pi)$ in the topological phase showing a band inversion for upper (u) and lower (l) branches with eigenvalues $\pm 1$ of $\hat{U}_{04}$. (c) Bulk energy gap versus $t'_z$ for $H$ with the topological phase (in green), with $t_z=2.5$. Arrow indicates parameters in Fig.~\ref{fig2}.}
    \label{fig3}
\end{figure}
We begin with the time-reversal-invariant Hamiltonians $\mathcal{H}(k_z=0)$ and $\mathcal{H}(k_z=\pi)$ that are effectively two-dimensional. For each effective Hamiltonian describing a finite system in the $xy$ plane, we divide the system into two regions $\{I, II\}$, where they share a common boundary with `twisted' hopping values - the original hoppings coupling the two regions are multiplied by a real-valued factor $\lambda$ that is continuously tuned from $1$ to $-1$, denoted by $\mathcal{H}_{\lambda}(k_z=0,\pi)$, see Fig.~\ref{fig3}(a). The choice of the cut and the twisted boundary condition ensure that the total system is time-reversal invariant, $C_2$-invariant as realized by the operator $\hat{U}_{04}$, and with two gauge-equivalent points $\mathcal{H}_{\lambda}$ and $\mathcal{H}_{-\lambda}$. Then, we define an unitary twist operator $\hat{V}_\lambda$ that satisfies $\hat{V}^\dag_\lambda \mathcal{H}_{\lambda} \hat{V}_\lambda = \mathcal{H}_{-\lambda}$. The twisted bulk-boundary correspondence utilizes the fact that the global $\hat{U}_{04}$ operator further anticommutes with the twist operator $\{\hat{V}_\lambda, \hat{U}_{04}\}=0$. With $\hat{U}_{04}$ having eigenvalues $\pm 1$, examining the evolution of the energy levels as a function of $\lambda$ can reveal two types of topology \cite{Song2020sicence}: either an adiabatic connection between states or, a spectral flow with gap closing as $\lambda$ continuosly changes from $1$ to $-1$, see Appendix \ref{A4}. 

For the three-dimensional quasicrystal we can locate the transition point between a trivial and a topological phase by examing the spectral flow of $\mathcal{H}_{\lambda}(k_z=\pi)$, see Fig.~\ref{fig3}(b). The transition takes place at the critical value $t'_z/t_z=0.6$, consistent with the bulk gap closing point and the concomitant proliferation of surface states above the critical value. A further increment to $t'_z/t_z=1.4$ closes the topological gap and brings the system back to the initial trivial insulator - this is due to a dimerization swapping in the $z$-direction [Fig.~\ref{fig3}(c)]. The robustness of the adapted twisted bulk-boundary correspondence can be checked for Fu's model of topological crystalline insulator \cite{Fu2011prl}, where it gives exactly the same phase diagram with the momentum space topology calculation.

\begin{figure}[t]
    \centering
    \includegraphics[width=0.48\textwidth]{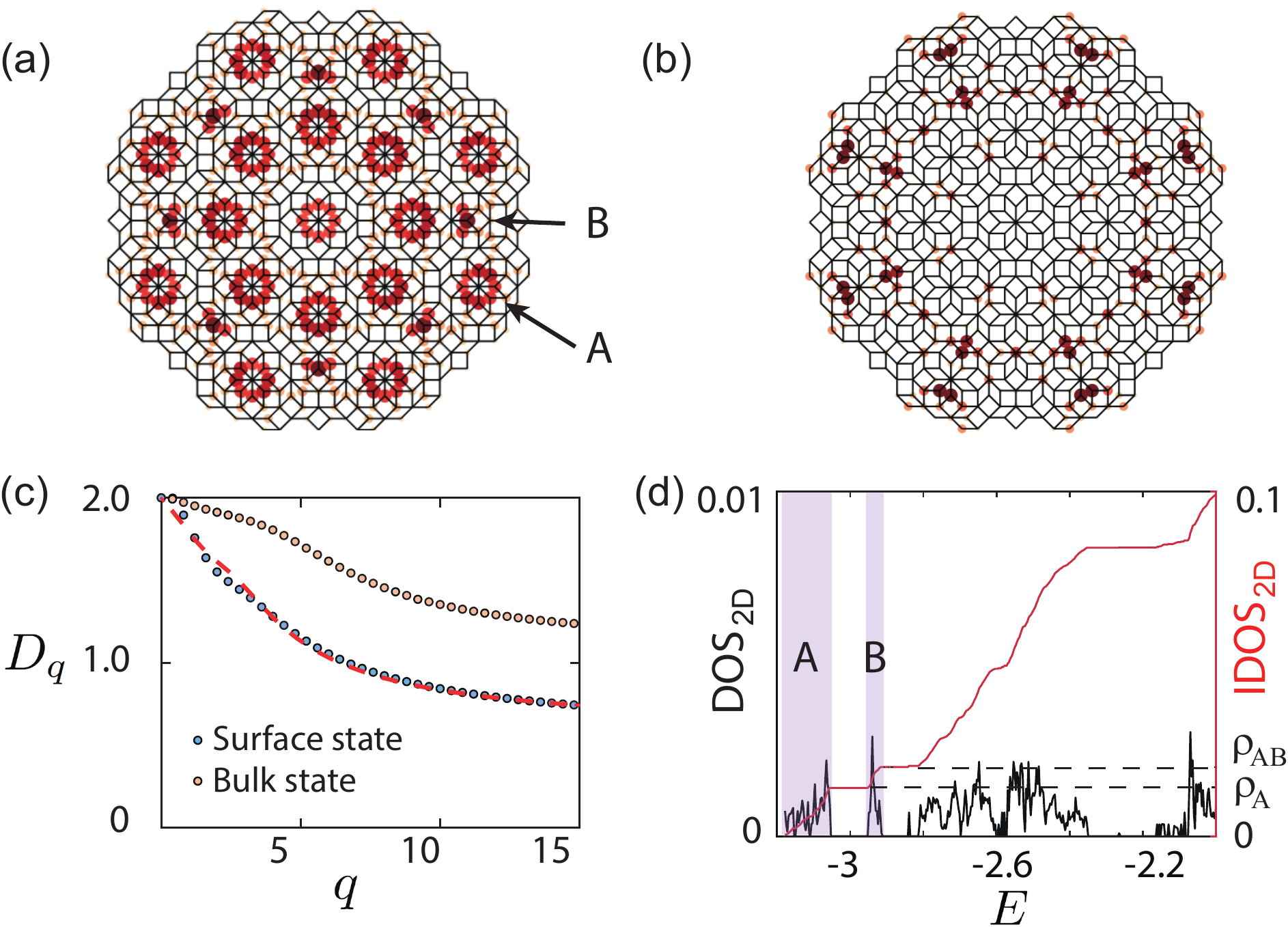}
    \caption{Fractal structure of eigenstates. The probability density distribution for (a) surface states with energy $E_\text{A}\pm0.05$ and with $689\times20\times2$ sites (b) a typical bulk state. (c) Anomalous scaling of moments from Eq. (\ref{eq4}). (d) Energy bands of the effective two-dimensional model with 23129 sites. Shaded band(s) composed of wavefunctions on A-(B-)patches. Dash red line in (c) are from the lowest band eigenstates.} 
    \label{fig4}
\end{figure}
\section{Fractal surface states and states counting}
To study the topological surface states structure on the quasicrytal plane, we define the in-plane eigenstate $\tilde{\phi}_E(\mathbf{r}_i)\equiv(\sum_{n,\sigma,\alpha}|\psi_E|^2)^{1/2}$ and evaluate its anomalous scaling behavior with respect to the planar size $N_r$:
\begin{equation}
P_q=\biggl\langle \sum_{\mathbf{r}_i}^{N_r}\tilde{\phi}_E(\mathbf{r}_i)^{2q}\biggr\rangle\sim N_r^{-\tau(q)/2}
\label{eq4}
\end{equation}
for large $N_r$. The average $\langle \cdot \rangle$ is performed with all eigenstates from a fixed energy shell $E_\text{A}\pm \delta$ surrounding the IPR peak, and the fractal dimension is given by $D_q= \tau(q) /(q-1)$. As shown in Fig.~\ref{fig4}(c), the variation of $D(q)$ as a function of $q$ signifies multifractality, where each moment scales anomalously with respect to the system size $N_r$. For comparison, multifractality to a varying degree is also obtained for typical bulk states [Fig.~\ref{fig4}(c)]. Thus, they both show characteristics of quasicrystal planar eigenstates and critical wavefunctions \cite{Kohmoto1987prb, Repetowicz1998prb, Jagannathan2017prb, Oktel2021prb, Evers2008rmp, Jagannathan2021rmp}. It must be emphasized that beside the main feature seen as concentrated covering of A-patches (and B-patches with a wider energy shell), the probability density has finite value beyond the covering zones to support multifractality. 

\begin{figure}[b]
    \centering
    \includegraphics[width=0.45\textwidth]{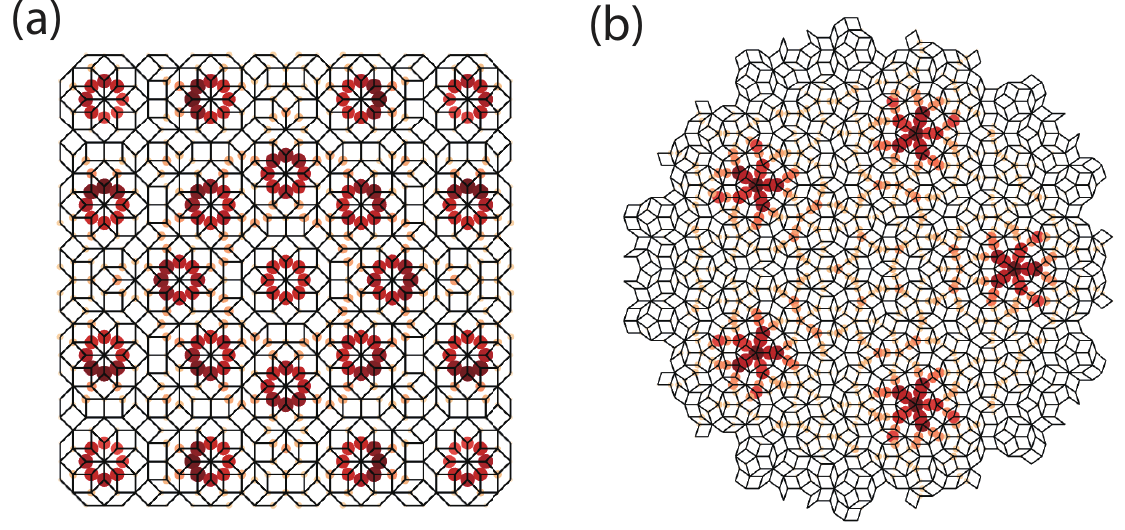}
    \caption{More general topological surface states. (a) Ammann-Beenker tiling with $C_4$ boundary (with $777\times 20\times2$ sites). (b) Penrose tiling with $C_5$ order, with $t_1=2.6$, $t_3=0.7$, $t_z'/t_z =0.8$, and $1381\times 9\times2$ sites.}
    \label{fig5} 
\end{figure} 
The structure of the surface states follows from considering the electron filling of an effective two-dimensional Ammann-Beenker tiling realized on the three-dimensional surface. By adopting a sufficiently strong two-dimensional nearest-neighbor hopping model, the integrated density-of-states ($\textrm{IDOS}_{\text{2D}}$) exhibits staircase structure giving rise to a simple `gap labelling' rule \cite{Vignolo2016prb, Bandres2016prx, Bellissard1992gap}, see Fig.~\ref{fig4}(d). Specifically, within this model, the two lowest plateaus result from two widely separated energy bands consisting solely of symmetric wavefunctions on A-patches and B-patches, respectively. A multifractal analysis for states from the lowest band gives exactly the same behavior as the topological surface states from the narrow energy shell [see red dash line in Fig.~\ref{fig4}(c)]. The plateau values, on the other hand, are simply determined by the fraction of sites belonging to A-patches and the totality of A- and B-patches, which are, in the thermodynamic limit, $\rho_{\textrm{A}}=(29-12\varphi)/2\approx 0.015$ and $\rho_{\textrm{AB}}=(-41+17\varphi)/2\approx 0.021$ where $\varphi=1+\sqrt{2}$ is the silver mean and the $1/2$ factor to account for two orbital per site (see Appendix \ref{A5}).

The topological surface modes in a finite sample can be understood as filling the two-dimensional electron band from the bottom, i.e., the number of near-degenerate surface states scales as $2\rho_{\textrm{A}} N_r$ at the IPR peak for a narrow energy shell with their probability density covering surface A-patches (and 2$\rho_{\textrm{AB}} N_r$ for a wider energy shell covering surface B-patches as well). A further widening of the energy shell results in a homogeneous real-space covering of surface sites, consistent with wavefunctions in the next-next-lowest energy band of the effective model which display a mixture of several patch types. Thus, the near-degenerate manifold of states, along with a regular covering of the surface, results in a macroscopic number of topological immobile states residing on the top (bottom) plane, which is in stark contrast to a crystalline topological surface band crossing. Furthermore, the localized nature of the surface states does not lead to quantized conductance (see Appendix \ref{A6}).
 
\section{Different boundary conditions and $C_5$ quasicrystal}
The robustness of the topological quasicrystal construction can be further addressed in two ways. First, we do not restrict the sample boundary in the $xy$ plane to respect the eigth-fold rotational order of Ammann-Beenker, see Fig.~\ref{fig5}(a). In the topological phase, we find A-patch surface modes as well as their scaling behavior continues to hold (see Appendix \ref{A2}). Second, by replacing the Ammann-Beenker tiling in Hamiltonian (\ref{eq1}) with Penrose tiling displaying a five-fold rotational order \cite{Senechal1996quasicrystals, baake2013aperiodic1}, we obtain the same topological phenomenon with surface states occupying the corresponding $S$-patches, see Fig.~\ref{fig5}(b). In this case, the combined orbital-real space rotation $\hat{U}_{ml}$ of Eq. (\ref{eq3}) is replaced by $C_{10}$ and $C_5$ generators, respectively.  It is also interesting to include spin-orbit effect in the model, which changes the topological class to the $Z_2$ topological insulator without crystalline symmetry \cite{Agarwala2017prl} (see Appendix \ref{A7}).

\section{Discussion}
Our proposal of three-dimensional topological quasicrystalline insulator is attractive in light of recent experimental realizations of simulating topological states including in photonics \cite{Bandres2016prx, Kim2022nc}, microwaves resonators \cite{Vignolo2016prb}, cold atoms \cite{Viebahn2019prl}, and acoustic metamaterials \cite{Peri2020sicence}. The model does not require spin-orbit coupling, which makes it feasible beyond the context of electronic systems. While quantum simulators can study eigenstate structure with energy resolution, i.e., focussing on surface state energies, state-of-the-art techniques of angle-resolved photoemission spectroscopy \cite{Lv2019np} should also allow for similar study of in-gap surface states below the Fermi energy in condensed matter systems.  Finally, the immobile surface modes with macroscopic near-degeneracy and time-reversal invariance may provide a novel platform for studying two-dimensional many-body effects. 

\begin{acknowledgments}
We thank Jean-No\"{e}l Fuchs, Zhao Liu, and Xin Wan for helpful discussions. This work is supported by NSFC project No. 11974308.
\end{acknowledgments}

\appendix
\section{Structure of Ammann-Beenker tiling}
\label{A1}

\begin{figure}[b]
	\centering
	\includegraphics[width=0.45\textwidth]{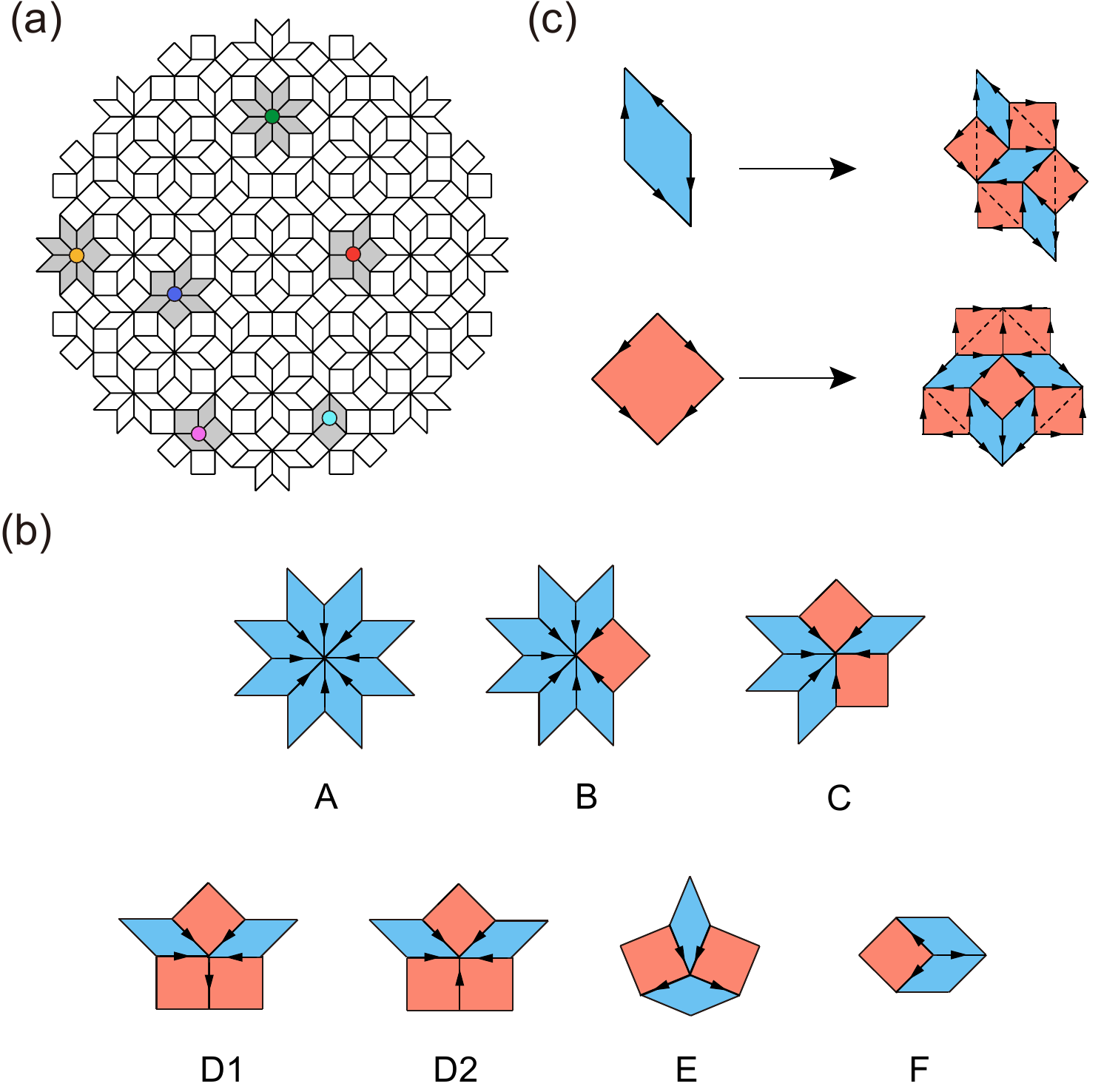}
	\caption{ (a) A sample of Ammann-Beenker tiling with 385 sites. The six patches are shaded with central site colored in green (A-), yellow (B-), blue (C-), red (D1/D2-), pink (E-) and cyan (F-patch), respectively. (b) The complete local patches with oriented links of Ammann-Beenker tiling. (c) Schematics of \textit{deflation-inflation method}: (1) Apply the subdivision rule to an initial local patch, such as an A-patch; (2) Inflate squares and rhombuses to their original sizes; (3) Repeat steps (1) and (2) $N$ times to obtain the tiling of different sizes.} 
	\label{figA1}
\end{figure}

An Ammann-Beenker tiling consists of square and rhombus (with the small angle $\pi/4$) elementary tilings, covering the full plane without mismatches or `defects', see Fig.~\ref{figA1}(a). The local tiling structure belongs to one of the seven types of local patch, conventionally known as A-, B-, C-, D1-, D2-, E-, and F-patch, respectively, referring to different local oriented connectivity emanating from each vertex [Fig.~\ref{figA1}(b)]. The tilings studied in this work are generated following the \textit{deflation-inflation method}~\cite{Senechal1996quasicrystals}, as illustrated in Fig.~\ref{figA1}(c). Note that the tiling shown in Fig.~\ref{figA1}(a) is generated from an initial A-patch, resulting in a finite sample which exhibits an eight-fold rotational invariance around the center.

\section{Surface states for different tiling sizes}
\label{A2}
\begin{figure}[h]
	\centering
	\includegraphics[width=0.45\textwidth]{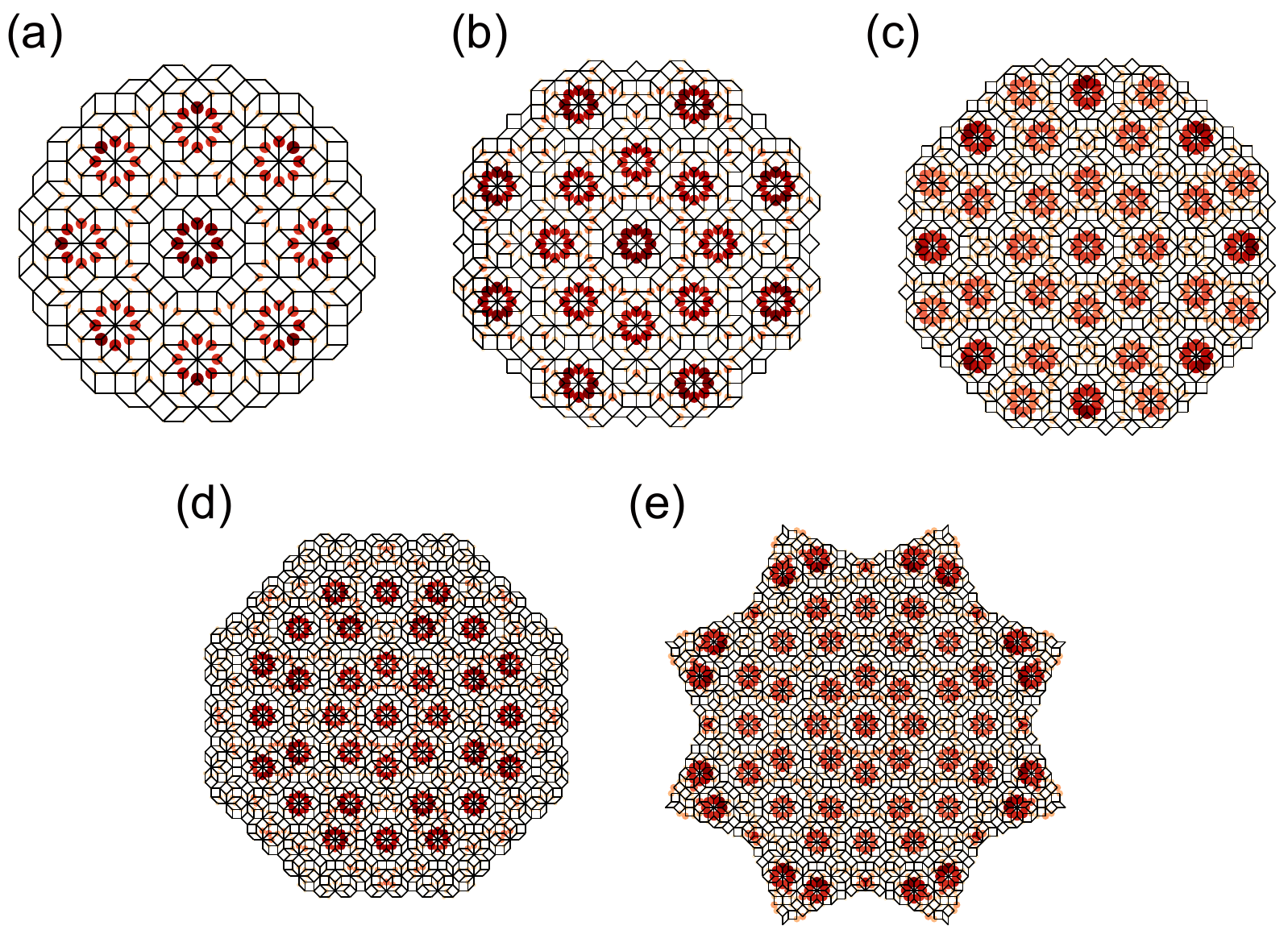}
	\caption{Probability density distributions of states within the narrow energy shell $E_{\text{A}}\pm0.03$ on the top-plane for various system sizes: (a) $329\times20\times2$; (b) $689\times20\times2$; (c) $1169\times14\times2$; (d) $1737\times9\times2$; (e) $1969\times9\times2$ sites indicating the planar size times the total number of bilayers.} 
	\label{figA2}
\end{figure}

The surface states within the narrow energy shell $E_{\text{A}}\pm\delta$ with $\delta=0.03$ are examined for various sizes of  the three-dimensional model with Hamiltonian (\ref{eq1}) in the main text. We find that the localization in the $z$-direction for the surface modes remains robust with increasing system size. In Fig.~\ref{figA2}(a)-(e), we show the evolution of the probability density distribution on the top-plane as the system size increases. Starting from (a) with planar size of 329 vertices, the near-degenerate manifold consists of nine states with three two-fold degenerate pairs and three isolated eigenstates close in energy. Among them, one state concentrates in the A-patch at the origin and the others concentrate in the remaining eight A-patches, forming the first generation of A-patch (see also Sec. III). By enlarging the planar size to 689 sites (b), another generation of A-patches appears in the wider perimeter, adding eight new near-degenerate states in the energy shell $E_{\text{A}}\pm\delta$. These behavior persists up to 1737 sites (c,d) as well as 1969 sites (e), with the latter having a more general boundary respecting the eight-fold rotational invariance, approaching the limit of computational capability. 

\section{Eigenstate structure from the multiplet}
\label{A3}
\begin{figure}[h]
	\centering
	\includegraphics[width=0.50\textwidth]{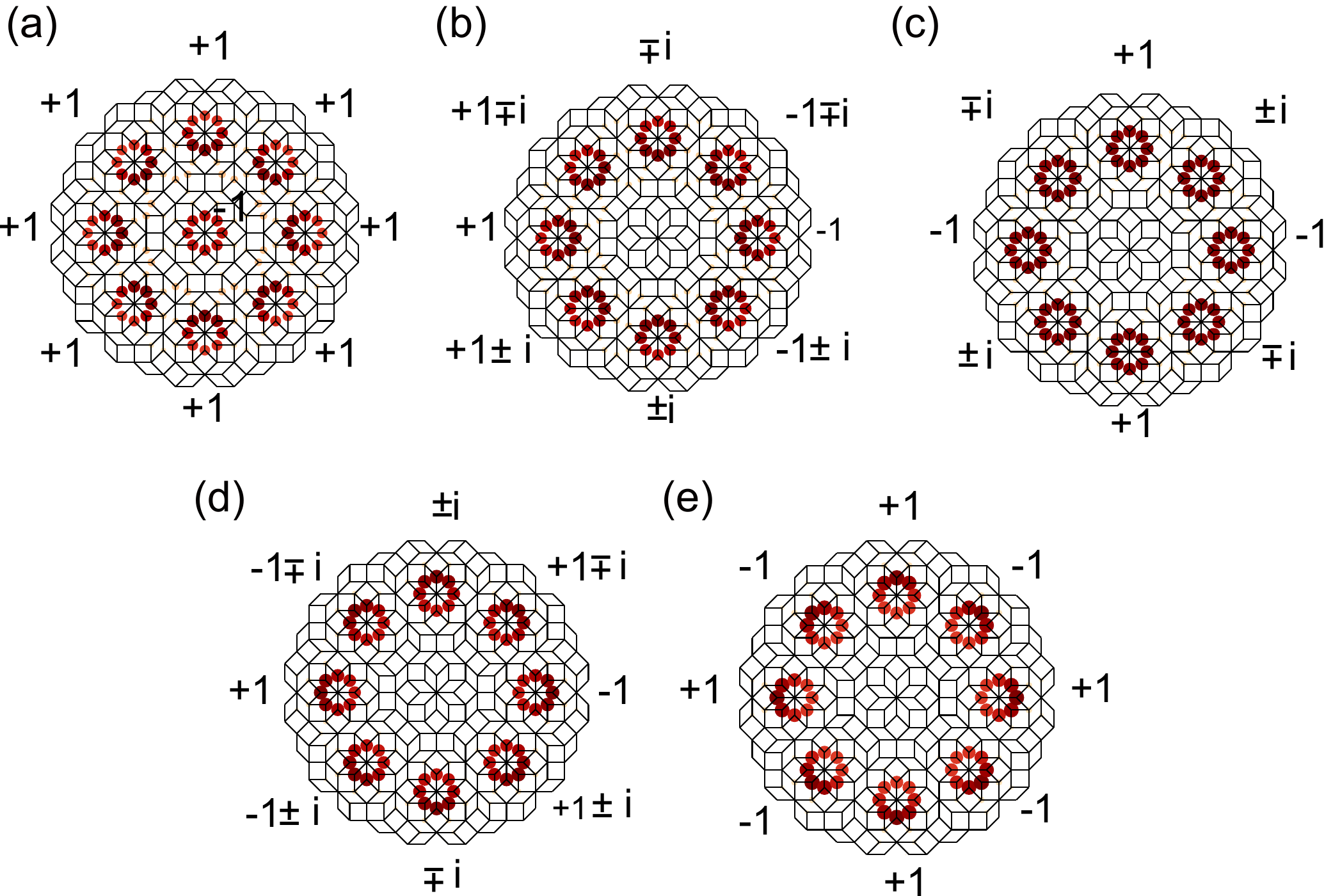}
	\caption{Eigenstates of $H_{\text{2D}}$ and $\hat{U}_{13}$ for Ammann-Beenker tiling of the lowest band. Complex amplitudes of the wavefunction are shown. The states are listed according to their eigenvalues of $\hat{U}_{13}$: (a) $+1$; (b) $-1\pm i$; (c) $\pm i$ ; (d) $1\pm i$; (e) $-1$, where (b), (c), (d) are separately two-fold energy degenerate.  Tight-binding parameters: $t_1=1.70$, $t_2=0.60$ and $t_3=0.15$ with 329 sites.}  
	\label{figA3}
\end{figure}

We plot the eight common eigenstates of $H_{\text{2D}}$ and $\hat{U}_{13}$ from the lowest band. Three pairs [Fig.~\ref{figA3}(b), (c) ,(d)] are strictly two-fold degenerate and combining the two other isolated eigenstates [Fig.~\ref{figA3}(a), (e)], they form a multipet in the near-degenerate manifold of states with probabilities concentrating on the first generation of A-patch.

\section{Spectral flow with twisted bulk-boundary correspondence}
\label{A4}
\begin{figure*}[t]
	\centering
	\includegraphics[width=0.75\textwidth]{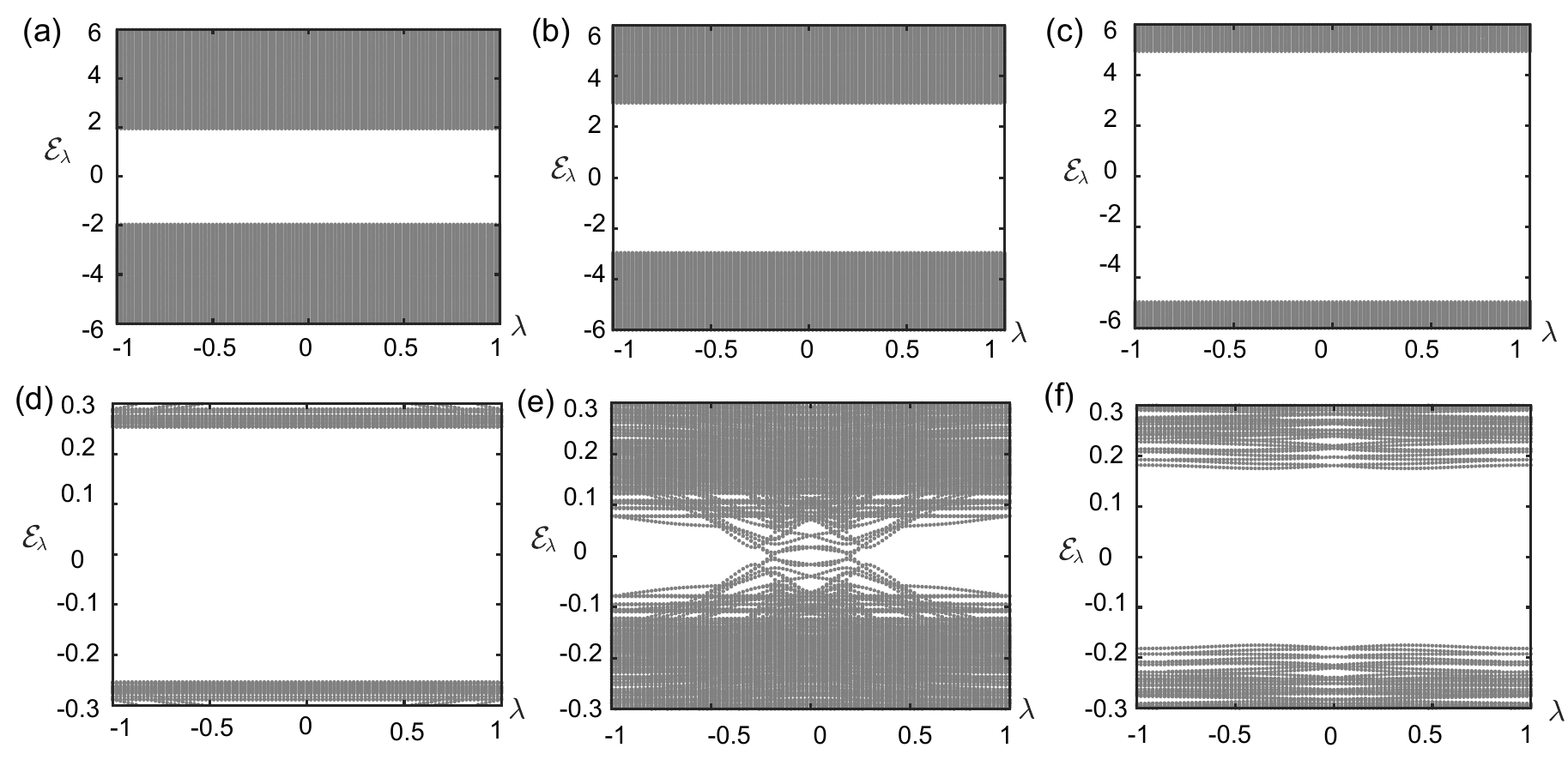}
	\caption{Determination of the topological phase transitions. (a)-(c) for $\mathcal{H}_{\lambda}(k_{z}=0)$ and (d)-(f) for $\mathcal{H}_{\lambda}(k_{z}=\pi)$ in various regimes. The first (a,d) and third (c,f) columns show no spectral flow for regimes $t'_z<1.5$ and $t'_z>3.5$. On the other hand, in (e) for the regime $1.5<t'_z<3.5$ it shows a spectral flow.}
	\label{figA4}
\end{figure*}

We numerically obtain the evolution of $\mathcal{E}_{\lambda}$ under the generalized twisted boundary condition \cite{Song2020sicence} in the three regimes of Fig.~3(c) in the main text, i.e., with different inter-bilayer coupling strengths $t_z'$. In Fig.~\ref{figA4}, the first and second row are for $\mathcal{H}_{\lambda}(k_{z}=0)$ and $\mathcal{H}_{\lambda}(k_{z}=\pi)$, respectively. Only in the middle panel with $1.5<t_z'<3.5$ in Fig.~\ref{figA4}(e) we obtain a spectral flow for $\mathcal{H}_{\lambda}(k_{z}=\pi)$ with band inversion (see also Fig.~3(b) in the main text where we label explicitly the band inversion). In the absence of a topological marker, we take this as an evidence of the topological phase transition in the three-dimensional quasicrystal, protected by quasicrytalline order and time-reversal invariance. The results are validated for $\mathcal{H}_{\lambda}(k_z)$ with Ammann-Beenker tiling up to 6080 sites, as well as Fu's topological crystalline insulator \cite{Fu2011prl}, showing the robust spectral flow.

\section{Gap labeling for Ammann-Beenker tiling}
\label{A5}
We give a derivation of the plateau values for the two lowest gaps in the IDOS$_{\text{2D}}$ for the effective $H_{\text{2D}}$ for Ammann-Beenker tiling discussed in the main text. 

We first define the ratio $p{(\xi)}=N_{\xi}/N_{\text{v}}$ as the fraction of each local patch in the full tiling, where $N_{\text{v}}$ is the total number of vertices and $N_{\xi}~(\xi=\text{A, B, C, D1, D2, E, F})$ represents the total number of each of the seven local patches [following Appendix \ref{A1}], respectively. We then employ a substitution matrix method as a generalization of that for Penrose tiling \cite{dePrunel2002prb, Vignolo2016prb, Bandres2016prx}, to solve for $p(\xi)$.  The so-called substitution matrix $S$ is constructed by taking nine elementary objects - square, rhombus and seven local patches (\text{A}, \text{B}, \text{C}, \text{D1}, \text{D2}, \text{E}, \text{F}), as a basis. The matrix elements of $S$ are obtained by following the subdivision rule in the \textit{deflation-inflation method} acting on these objects, see caption in Fig.~\ref{figA1}(c). The first column of $S$, for example, follow from the subdivision of a square into three squares, four rhombuses, one $\text{D1}$-patch, two $\text{E}$-patches and three $\text{F}$-patches. The full $S$ matrix is given by
\begin{equation}
	S=\begin{pmatrix}
		3 & 2 & 0 & 0 & 0 & 0 & 0 & 0 & 0 \\
		4 & 3 & 0 & 0 & 0 & 0 & 0 & 0 & 0 \\
		0 & 0 & 1 & 1 & 1 & 0 & 1 & 0 & 0 \\
		0 & 0 & 0 & 0 & 0 & 1 & 0 & 0 & 0 \\
		0 & 0 & 0 & 0 & 0 & 0 & 0 & 1 & 0 \\
		1 & 0 & 0 & 0 & 0 & 0 & 0 & 0 & 0 \\
		0 & 0 & 0 & 0 & 0 & 0 & 0 & 0 & 1 \\
		2 & 2 & 0 & 0 & 0 & 0 & 0 & 0 & 0 \\
		3 & 2 & 0 & 0 & 0 & 0 & 0 & 0 & 0
	\end{pmatrix}.
\end{equation}
The largest eigenvalue of $S$ is $\varphi^2$ (with silver mean $\varphi=1+\sqrt{2}$) and components of the corresponding normalized eigenvector determine ratios for the nine objects, respectively. Specifically, for the seven local patches they are determined numerically to give 
\begin{equation}
\begin{aligned}
	&(p(\text{A}),p(\text{B}),p(\text{C}),p(\text{D1}),p(\text{D2}),p(\text{E}),p(\text{F}))=\\
	&(\varphi^{-4},\varphi^{-5},2\varphi^{-4},\varphi^{-3},\varphi^{-3},2\varphi^{-2},\varphi^{-1}).
\end{aligned}
\end{equation}
Relevant to fixing the first and second plateaus in the IDOS$_{\text{2D}}$ are $p(\text{A})$ and $p(\text{B})$ since the lowest and second-lowest bands of the two-dimensional model consist solely of wavefunctions concentrating on A-patch (octamer-) and B-patch (heptamer-like structure), respectively. We thus arrive at 
\begin{equation}
	\rho_{\text{A}}=p(\text{A})/2=(29-12\varphi)/2\approx0.015,
\end{equation}
\begin{equation}
	\begin{aligned}
		\rho_{\text{AB}}=(p(\text{A})+p(\text{B}))/2=(-41+17\varphi)/2\approx0.021.
	\end{aligned}
\end{equation}
using $\varphi^{-4}=29-12\varphi$, $\varphi^{-5}=-70+29\varphi$, and the factor $1/2$ is to account for two orbitals per site. The result written in the form $\rho_{i}=\mathbb{Z}_{i_{1}}+\mathbb{Z}_{i_{2}}\varphi$, with $i$ the band index and $\mathbb{Z}_{i_{1,2}}$ are relative integers, is known in the literature as a gap labeling theorem \cite{Vignolo2016prb, Bandres2016prx, Bellissard1992gap}.\\

\section{Simulation of surface state transport}
\label{A6}
\begin{figure}[h]
	\centering
	\includegraphics[width=0.45\textwidth]{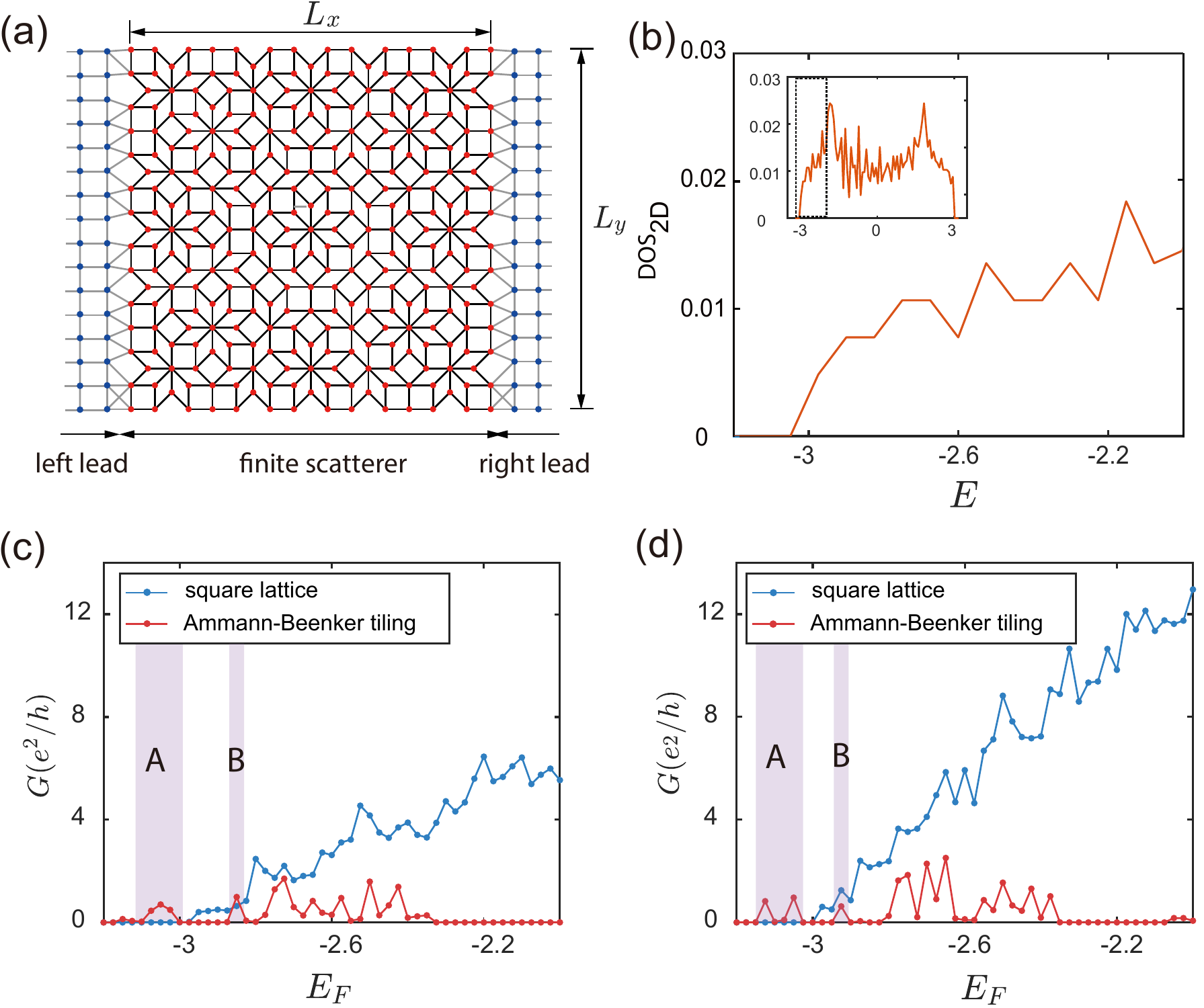}
	\caption{(a) Schematic of the setup with Ammann-Beenker tiling of size $L_x \times L_y$. (b) For comparison, we show the density-of-states ($\textrm{DOS}_{\textrm{2D}}$) for a square lattice with $L_x=L_y=32$ (Inset: DOS in full energy range). Two-terminal conductance (red: Ammann-Beenker; blue: square lattice) with two sample sizes: (c) $L_x=L_y=16$; (d) $L_x=L_y=32$. Shaded $A$ and $B$ regions correspond to the energy bands for eigenstates with A- and B-patch localization (\textit{cf}. Fig. 4).}   
	\label{figA5}
\end{figure}

To model the transport property of the topological surface state, we use the two-dimensional tight-binding Hamiltonian of Ammann-Beenker quasicrystal and compute its two-terminal conductance via the Landauer-Büttiker method \cite{Datta1997electronic,Datta2005quantum}. For the conductance $G=\frac{e^2}{h}T(E_F)$, where $T$ is the total transmission coefficient, we tune the Fermi energy $E_F$ of the effective model from the bottom to the next lowest band, to simulate the transport stemming from A- and B-patch eigenstates of the surface states, summarized in Fig.~\ref{figA5}. 

We compute the conductance with two sample sizes (red in Fig.~\ref{figA5}(c), (d)), and correspondingly the square lattice case (blue). We see that the localized nature of the eigenstates renders the system a bad metal, which can be contrasted with the metallic case of the square lattice. By doubling the system size, we see a proportional increase in the conductance for the metal, whereas for the bad metal it does not. Furthermore, the dips in the conductance are due to the presence of mini-gaps in the DOS. The analysis confirms that the multifractal A- and B-patch states exhibit a unique kind of response intermediate between an insulator and a metal.

\section{Strong $Z_2$ topological insulator with quasicrystal}
\label{A7}

The three-dimensional strong $Z_2$ topological insulator can be constructed by stacking Ammann-Beenker quasicrystals and including spin-orbit effect. We follow the amorphous Hamiltonian proposed in Ref. \cite{Agarwala2017prl} with the hopping matrix $T_{\alpha\beta}(\theta,\phi)$ between two different sites, within a $\sqrt{2}$-distance of the edge length: 

\begin{equation}
\begin{aligned}
		&T_{\alpha\beta}(\theta,\phi)=\\
		&\frac{1}{2}
\begin{pmatrix}
	1 & 0 & -i\cos\theta & -ie^{-i\phi}\sin\theta \\
	0 & 1 & -ie^{i\phi}\sin\theta & i\cos\theta \\
	-i\cos\theta & -ie^{-i\phi}\sin\theta & -1 & 0 \\
	-ie^{i\phi}\sin\theta & i\cos\theta & 0 & -1 \\
\end{pmatrix},
\end{aligned}
\end{equation}
where $\alpha,\beta$ are indices for the spin-orbital space, $\phi$ and $\theta$ are the in-plane azimuthal and polar angles between two sites, respectively. In addition, there is the on-site potential $\epsilon_{\alpha\beta}$=Diag$(-3+M,-3+M,3-M,3-M)$ with mass $M$.

\begin{figure}[h]
	\centering
	\includegraphics[width=0.48\textwidth]{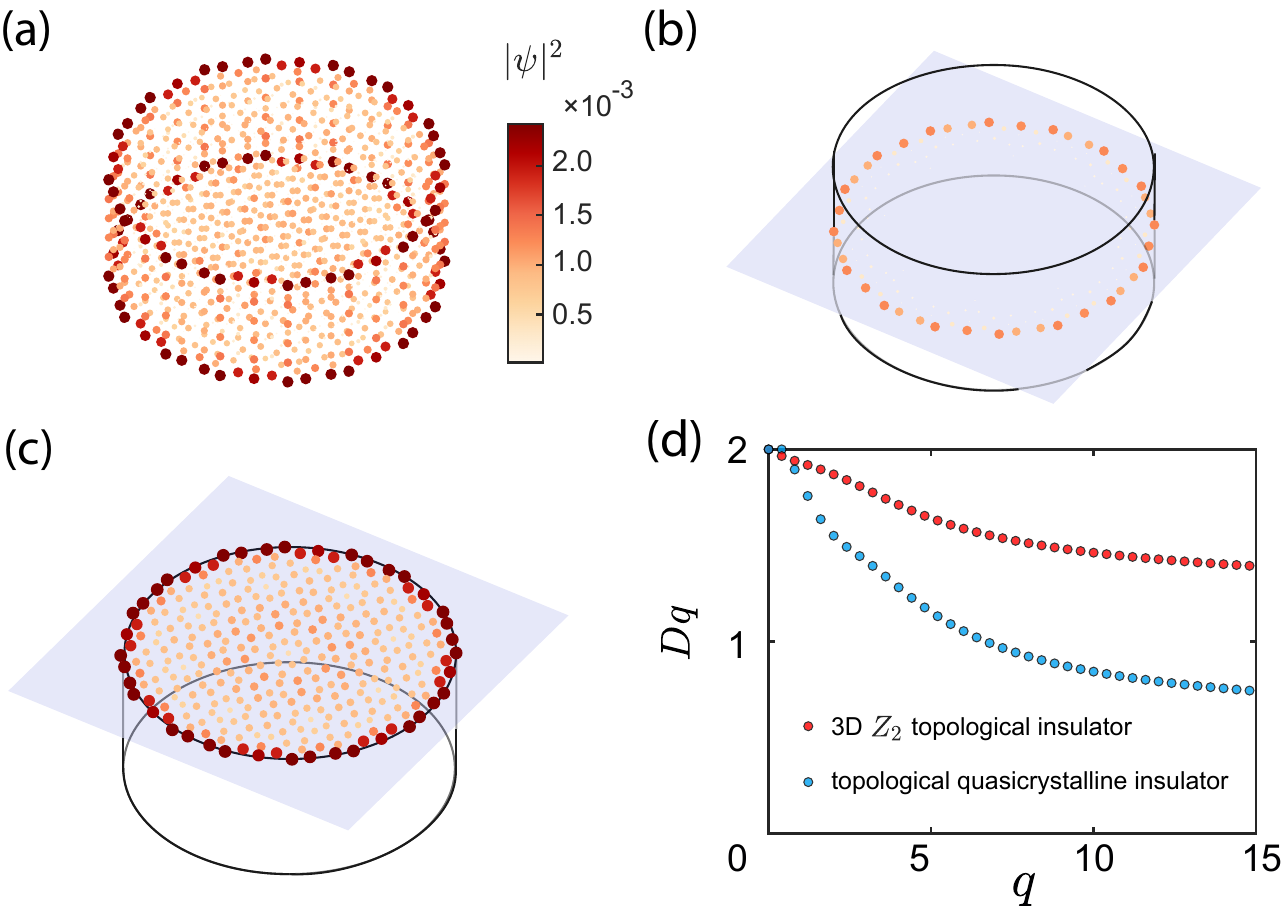}
	\caption{(a) Probability density distribution of the surface state in the three-dimensional strong $Z_2$ topological model with $M=0.5$ (system size: 329×6×4). Cut across the center (b) and the top surface (c) of the system. (d) Multifractal analysis of the surface states (red) with energy shell $E=0\pm0.2$. Blue dots are for the topological quasicrystalline insulator surface states.}   
	\label{figA6}
\end{figure}

The topological surface state is shown in Fig.~\ref{figA6}(a)-(c), showing a very different characteristics from the topological quasicrystalline insulator in the main text, \textit{cf}. Fig.~\ref{fig1}(b). Specifically, it resides on all surfaces rather than only on certain surfaces of the system, reflecting the strong topological nature of the phase. Moreover, in comparison, the probability density does not show concentration in high symmetry patches. This is further evidence in the the multifractal analysis shown in Fig.~\ref{figA6}(d), in red, with the scaling of moments similar to the fractal tiling itself. This supports the notion that topological quasicrystalline insulator studied in the main text belongs to a different class than that of the strong $Z_2$.

\end{document}